\documentclass[%
preprint,
superscriptaddress,
amsmath,amssymb,
prl,
]{revtex4-2}
\usepackage[T1]{fontenc}
\usepackage{color}
\usepackage{xcolor}
\usepackage{soul}
\usepackage{float}
\usepackage[pdftex]{graphicx}
\usepackage{graphicx}
\usepackage{dcolumn}
\usepackage{bm}
\usepackage{siunitx}
\usepackage{comment}
\usepackage{fixltx2e}
\usepackage{soul}
\usepackage{braket}
\usepackage{lineno}
\usepackage{siunitx}
\usepackage{afterpage}

\usepackage{hyperref}
\hypersetup{
    bookmarks=true,         
    unicode=false,          
    pdftoolbar=true,        
    pdfmenubar=true,        
    pdffitwindow=false,     
    pdftitle={A foundry-fabricated spin qubit unit cell with in-situ dispersive readout},    
    pdfauthor={Dr. Matias Urdampilleta},     
    pdfsubject={},   
    pdfcreator={Dr. Matias Urdampilleta},   
    pdfproducer={},  
    pdfkeywords={,} {} {}, 
    pdfnewwindow=true,      
    colorlinks=false,       
    linkcolor=red,          
    citecolor=green,        
    filecolor=magenta,      
    urlcolor=cyan           
}

\usepackage{filecontents}
\usepackage{comment}

\DeclareUnicodeCharacter{2009}{\,} 
\DeclareSIUnit{\belmilliwatt}{Bm}
\DeclareSIUnit{\dBm}{\deci\belmilliwatt}

\newcommand{\tss}[1]{\textsubscript{#1}}
\newcommand{\updown}{$\vert\uparrow\downarrow\rangle\; $} 
\begin{document}
\bibliographystyle{apsrev4-1}
\setstcolor{red}
\preprint{APS}
\title{A foundry-fabricated spin qubit unit cell with in-situ dispersive readout}

\author{Pierre Hamonic}
\affiliation{Univ. Grenoble Alpes, CNRS, Grenoble INP, Institut N\'eel, 38402 Grenoble, France}

\author{Mathieu Toubeix}
\affiliation{Univ. Grenoble Alpes, CEA, List, F-38000 Grenoble, France}

\author{Guillermo Haas}
\affiliation{Univ. Grenoble Alpes, CNRS, Grenoble INP, Institut N\'eel, 38402 Grenoble, France}

\author{Jayshankar Nath}
\affiliation{Quobly, Grenoble, France}

\author{Matthieu C. Dartiailh}
\affiliation{Quobly, Grenoble, France}

\author{Biel Martinez}
\affiliation{Univ. Grenoble Alpes, CEA, Leti, F-38000 Grenoble, France}

\author{Benoit Bertrand}
\affiliation{Univ. Grenoble Alpes, CEA, Leti, F-38000 Grenoble, France}

\author{Heimanu Niebojewski}
\affiliation{Univ. Grenoble Alpes, CEA, Leti, F-38000 Grenoble, France}

\author{Maud Vinet}
\affiliation{Quobly, Grenoble, France}

\author{Christopher B{\"a}uerle}
\affiliation{Univ. Grenoble Alpes, CNRS, Grenoble INP, Institut N\'eel, 38402 Grenoble, France}

\author{Franck Balestro}
\affiliation{Univ. Grenoble Alpes, CNRS, Grenoble INP, Institut N\'eel, 38402 Grenoble, France}

\author{Tristan Meunier}
\affiliation{Univ. Grenoble Alpes, CNRS, Grenoble INP, Institut N\'eel, 38402 Grenoble, France}
\affiliation{Quobly, Grenoble, France}

\author{Matias Urdampilleta	}
\email{matias.urdampilleta@neel.cnrs.fr}
\affiliation{Univ. Grenoble Alpes, CNRS, Grenoble INP, Institut N\'eel, 38402 Grenoble, France}

\date{\today}

\begin{abstract}
Spin qubits based on semiconductor quantum dots are a promising prospect for quantum computation because of their high coherence times and gate fidelities. However, scaling up those structures to the numbers required by fault-tolerant quantum computing is currently hampered by a number of issues. One of the main issues is the need for single-shot low-footprint qubit readout.
Here, we demonstrate the single-shot in situ measurement of a compact qubit unit-cell. The unit cell is composed of two electron spins with a controllable exchange interaction. We report initialization, single-shot readout and two-electron entangling gate. 
The unit cell was successfully operated at up to 1 K, with state-of-the-art charge noise levels extracted using free induction decay. With its integrated readout and high stability, this foundry fabricated qubit unit cell demonstrates strong potential for scalable quantum computing architectures.

\end{abstract}

\maketitle

\section{INTRODUCTION}
Silicon spin qubits have a strong potential for scalability due to their compacity and compatibility with industrial CMOS processes\cite{Watson2018, Philips_2022, Huang2024, Takeda2022, Weinstein2023, Thorvaldson2024, Chatterjee2021}. As a result, the transition from academic fabrication to industrial manufacturing is expected to lead to drastic improvements in the reliability and quality of devices \cite{ Zwerver2022, Crawford2023, Weinstein2023, Neyens2024, Thomas2025, 10639218}. 
Growing efforts in this direction have led to recent progress towards producing high-quality two-qubit gates in 300 mm foundry-fabricated unit cells \cite{Steinacker2024}. However, the need for charge sensor-based qubit readout systems requires additional electrostatic tuning and hinders scale-up as a unit cell \cite{Veldhorst2017}.

In this context, in situ qubit readout using gate-based dispersive methods could be a viable approach \cite{jung2012radio, house2016high, hornibrook2014frequency}, especially if the high electrostatic coupling offered by industrial gate stacks is exploited \cite{Oakes_2022, PhysRevX.5.031024, gonzalez2015gate, crippa2019}.
This technique was specifically used to discriminate a singlet from a triplet state in a large magnetic field, and its discriminatory capacity has been widely used for the single-shot characterization of spin relaxation and related spectra \cite{Zheng2019, Pakkiam2018, DzurakReflecto}. 
However, single-shot gate-based readout on a qubit remains to be demonstrated.

In this paper, we demonstrate the in situ readout of a foundry-fabricated spin qubit unit cell. Our approach eliminates the need for an external charge sensor, producing a compact unit cell with a minimal number of electrostatic controls.
A single-shot readout from a coherently driven singlet-triplet qubit shows good visibility. Building on this readout, characterization of the qubit reveals high-fidelity initialization, high-frequency singlet-triplet rotation, and very low charge noise levels linked to low-frequency qubit energy fluctuations.
Finally, we demonstrate that the qubit has good quality metrics and reasonable visibility up to 1 K, demonstrating strong potential for cointegration with dissipative cryoelectronics \cite{Ruffino_2021, Xue_2021, Bartee2024}.

\section{The spin qubit unit cell}

\begin{figure}%
\includegraphics[width=\columnwidth]{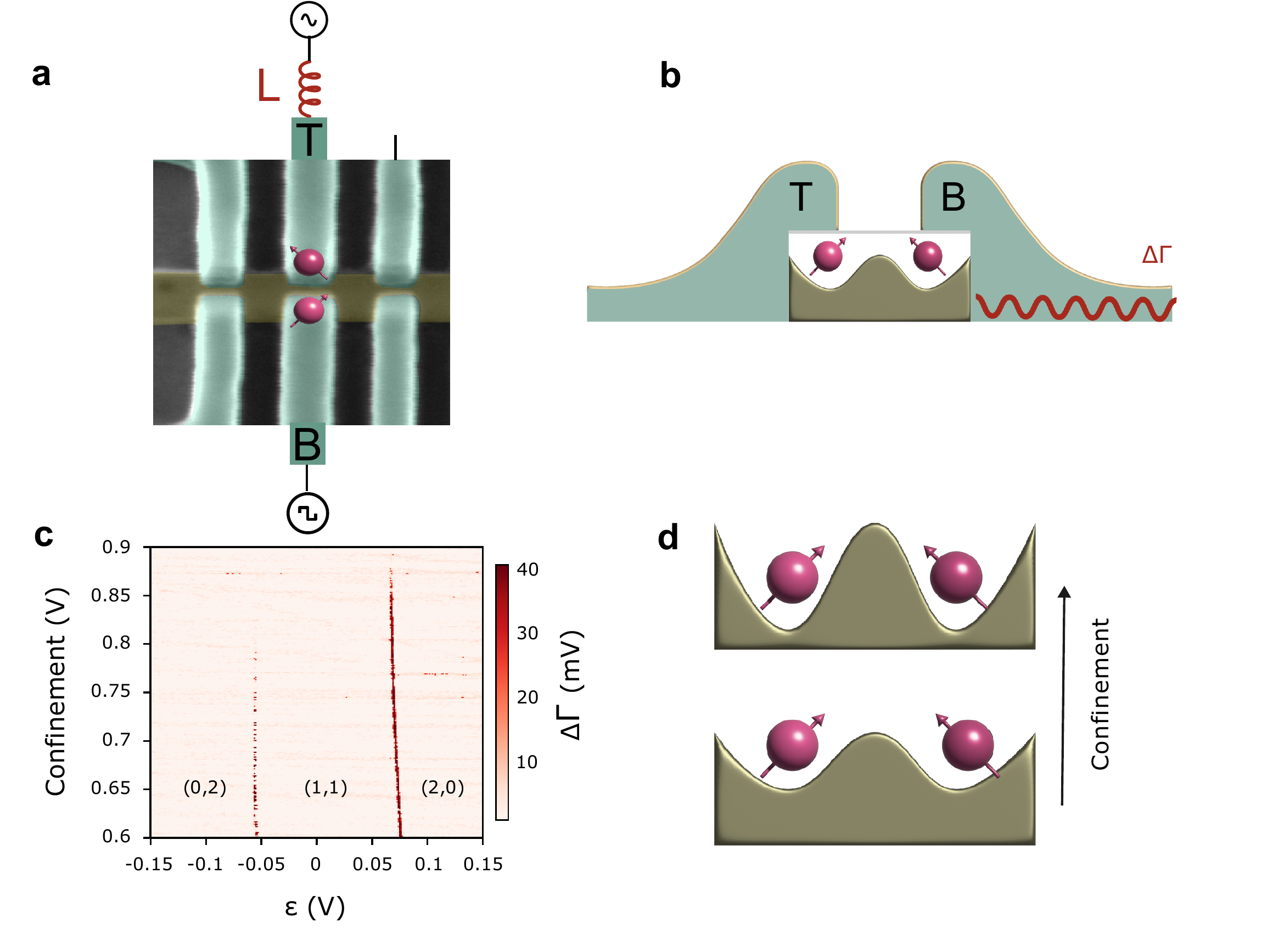}
\caption{
\textbf{The foundry fabricated spin qubit unit cell} 
\textbf{(a)} Scanning electron microscopy (SEM) of the split-gate device with false colors. The silicon channel (khaki) is covered by split gates (pastel green) with a SiO$_2$/TiN/polysilicon gatestack.
A double quantum dot is  formed under gates T and B and loaded with a total of 2 electrons. 
The T gate is connected to a tank circuit formed by the inductance L and the device's parasitic capacitance to ground. 
The B gate is  connected to a bias tee to allow application of rapid pulses to electrically manipulate the singlet-triplet qubit.
\textbf{(b)} Artistic representation of a cross-section of the device showing the isolated double dot formed between the two split-gates.
\textbf{(c)} Charge stability diagram obtained by measuring the shift in reflected signal $\Delta\Gamma$ on T as the detuning $\varepsilon$ and the confinement potential C are varied. The two inter-charge transitions delimiting the three charge regions are clearly visible.
\textbf{(d)} Artistic representation of how confinement potential affects the height of the tunnel barrier between the two quantum dots.
}
\label{fig:Device}%
\end{figure}

This study uses the foundry-fabricated device described by ref \cite{Hamonic2024} (Fig.\ref{fig:Device} (a)). 
The device is patterned on a 300 mm wafer in the CEA-Leti industrial research foundry using fully depleted silicon on insulator (FDSOI) technology.
It features a 40 nm wide 10 nm thick silicon channel. 
The two extremities of the channel, labeled source (S) and drain (D), are n-doped to create reservoirs of electrons; they are separated from the substrate by a 145-nm layer of buried oxide. 
Titanium nitride and poly-silicon gates are deposited on top of the channel, over a 6-nm isolation layer of thermally-grown dioxide.
A combination of deep-ultraviolet and electron-beam lithography is used to pattern the gate structure, as described in\cite{PRXQuantum.3.040335, Klemt2023}.

All gates are 40 nm wide and spaced 40 nm apart in both the longitudinal and transverse directions.
To prevent unwanted doping of the channel, silicon nitride spacers (35 nm thick) are deposited between the gates before doping the reservoirs.
Finally, the whole device is encapsulated and the gates are connected to aluminum bond pads through standard Cu-damascene back-end-of-line processing.

The unit cell used here contains two gates labeled T and B, for top and bottom, in reference to their position with respect to the silicon channel, see Fig. \ref{fig:Device}(a,b) \cite{PhysRevApplied.14.024066}.
T is connected to a superconductive inductor with $L = \SI{69}{nH}$, together with the parasitic capacitance $C_{p} = \SI{0.25}{\pico\farad}$, it forms a resonator with resonance frequency $f = \SI{1148}{\mega\hertz}$ and quality factor around 50.
The inductor is further connected to a bias tee with cutoff frequency $f\tss{c} = \SI{100}{\kilo\hertz}$ to allow application of both a reflectometry tone and a DC voltage. The amplitude variation ($\Delta\Gamma$ in Fig. \ref{fig:Device}(b,c)) of the reflected radio-frequency signal close to the resonance frequency is determined using analog demodulation at room temperature.
B is also connected to a bias tee with the same cutoff frequency, this time to allow fast manipulation pulses to be applied together with a DC voltage.

The other gates are DC-biased and only relevant during the loading sequence and for isolation of the unit cell.

At the 80 mK base temperature of the dilution fridge, when a positive voltage is applied to the gates, charges accumulate at the Si/SiO\tss{2} interface, forming quantum dots.
We can load a finite number of electrons into T and B and then isolate a double quantum dot (DQD) from the reservoirs following our published procedure  \cite{Hamonic2024}.
Hereafter, we work with two electrons in the isolated DQD \cite{Bertrand2015}, there are thus three possible charge states in the isolated regime: (0,2), (1,1), and (2,0).

To tune the DQD, we define detuning $\varepsilon = \frac{V_T - V_B}{2}$ and the confinement potential $C = \frac{V_B + V_T}{2}$, as both gates have similar lever arm ratio.
Fig \ref{fig:Device} (c) shows the resonator response as $\varepsilon$ and C vary.
Two parallel interdot charge transitions (ICT) are observed, separating the three possible charge states. 

These ICTs tend to disappear toward high potential depth. We interpret this effect as a result of a drop off in tunnel coupling, to the point that the transitions are no longer detectable. Indeed, a large potential depth corresponds to strong confinement, which creates a larger barrier between the dots and thus necessarily reduces tunnel coupling, see Fig. \ref{fig:Device}(d) \cite{ivlev2025operating}. 
We exploit this confinement to fine-tune the tunnel coupling and the exchange interaction between two electrons in our device.

\section{Single-shot readout and initialization}

\begin{figure*}%
\includegraphics[width=\columnwidth]{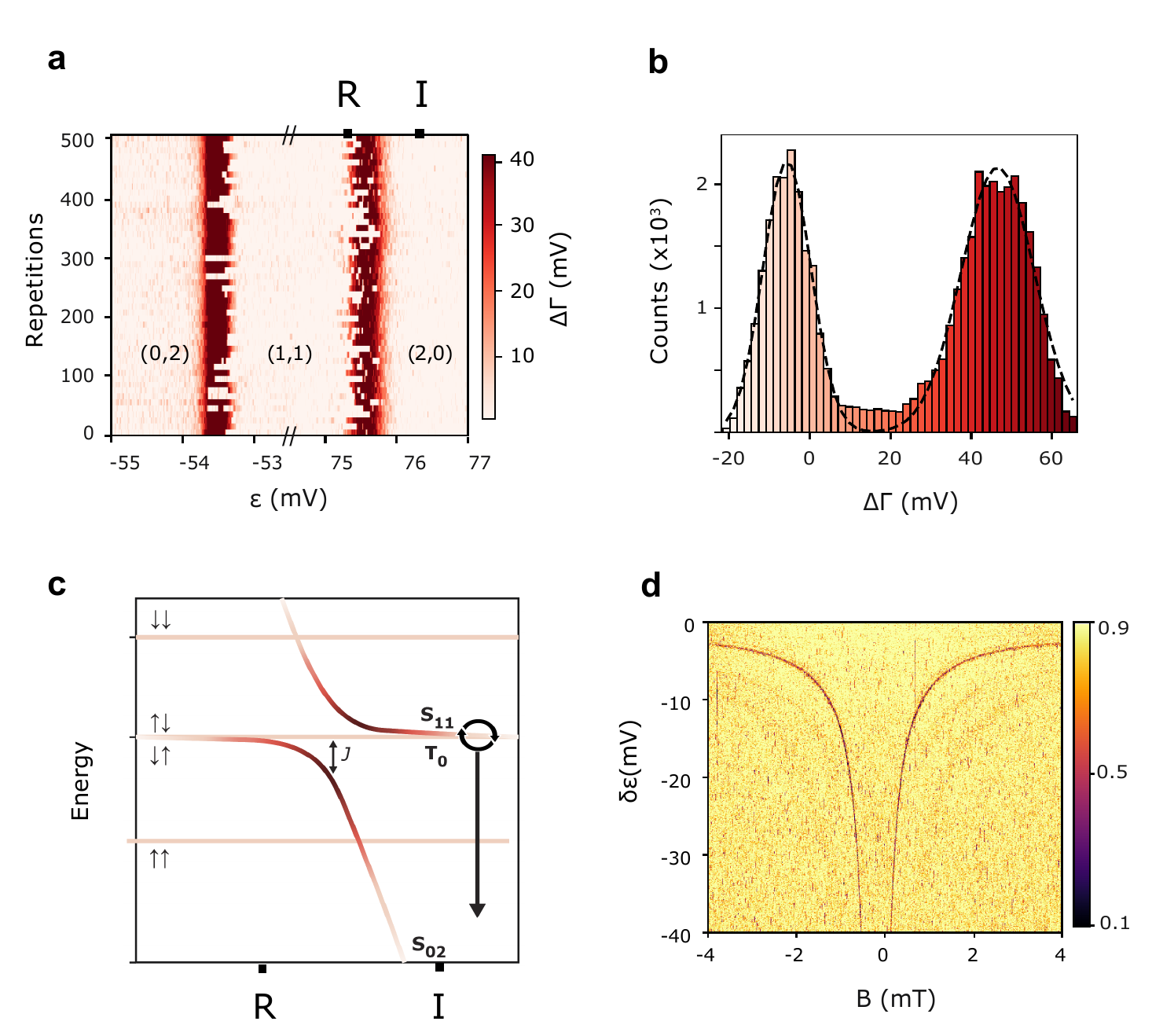}
\caption{
\textbf{Single-shot readout and initialization.} 
\textbf{(a)} Horizontal cross-section of Fig. \ref{fig:Device}(c) taken at Confinement C = 0.7 V for 500 repetitions. 
On both ICTs, (0,2)$\leftrightarrow$(1,1) and (1,1)$\leftrightarrow$(2,0), thermal singlet to triplet excitations combine with PSB to produce pixels with zero signal. 
\textbf{(b)} Histogram obtained by binning the repeated signal at measurement position R on the (1,1)$\leftrightarrow$(2,0) transition to probe the singlet and triplet  populations.
\textbf{(c)} Energy diagram for the qubit, highlighting the relevant states (S\tss{11}, T\tss{0}, and the two polarized triplets) and the positions for initialization (I) and readout (R).
\textbf{(d)} Spin funnel experiment showing the S-T\textsuperscript{-} anticrossing as a function of magnetic field and detuning variation. From the spliting at zero-field, we extract the residual exchange interaction at zero detuning (deep in (1,1)), $J\tss{0} = \SI{60}{\mega\hertz}$. The field is slightly offset due to hysteresis in the magnetic coil.}
\label{fig:Readout}%
\end{figure*}

We read our qubit by measuring the difference in reflection of a radio-frequency tone on the tank circuit at a frequency close to the resonance frequency $f = \SI{1148}{\mega\hertz}$.
The resonator is designed to work in the adiabatic regime, where the probe frequency is smaller than the tunnel coupling in the standard spin qubit operating regime  (t\tss{c} \textless  \SI{5}{\giga\hertz}).
In this regime, a small excitation is used to probe the dispersion relation of the quantum states.
Therefore, when the detuning energy approaches an interdot charge transition, an additional quantum capacitance C\tss{Q} arises due to the tunnel coupling \cite{Vigneau2023}. 
This alters the resonance frequency of the resonator and, consequently, the reflected signal. 
In our device, this shift, $\Delta\Gamma$, is measured through homodyne detection.

This detection scheme enables spin readout of the unit cell as follows: due to Pauli spin blockade (PSB), only singlet states can readily tunnel from (1,1) to (0,2) or (2,0) and give finite quantum capacitance. 
In contrast, triplet states are blocked, and thus present zero quantum capacitance, leaving the RF signal unchanged. Singlet and triplet states can therefore be readily distinguished at the ICT.

PSB manifests as stochastic signal jumps in the ICT diagram. 
As detuning is scanned across the ICTs  (1,1)$\leftrightarrow$(0,2), and (1,1)$\leftrightarrow$(2,0) (see Fig. \ref{fig:Readout}(a)), the finite and zero quantum capacitance are represented as red and blank pixels, respectively. 
We attribute the fluctuations observed to thermally excited triplet states that relax at a rate comparable to the measurement time (1 ms/pixel).

The PSB signature is the same for both transitions in Fig. \ref{fig:Readout}(a)). Consequently, although in the rest of this article, results for the (1,1)$\leftrightarrow$(2,0) ICT are presented, the reader should note that similar results were obtained for the (0,2)$\leftrightarrow$(1,1) transition (see supplementary materials).

To characterize the single-shot readout fidelity, we exploit PSB at the ICT.
We place our measurement position at the point of maximum contrast on the (1,1)$\leftrightarrow$(2,0) transition and record 10 000 samples. The data are illustrated in the histogram displayed in Fig. \ref{fig:Readout}(b). Fitting the histogram with two bell curves and adjusting the threshold to maximize visibility, we obtain a charge fidelity of 97\% for an integration time of \SI{100}{\micro\second}.
We use the same method hereafter to define the threshold and extract single-shot measurements. 

We next apply the readout method as part of a spin-funnel experiment \cite{Fogarty_2018}.
We start by initializing a singlet state by spending time ($t = \SI{100}{\micro\second}$) at a relaxation hotspot in (2,0), labelled I in Fig. \ref{fig:Readout}(a,c), until all triplets have relaxed. 
At low field, the triplet relaxation rate is strongly enhanced by spin state mixing \cite{PRXQuantum.4.010329} due to spin-orbit coupling \cite{PhysRevX.9.021028}, or residual hyperfine interactions. 
By monitoring these parameters, we determine a greater than 95\% initialization fidelity in the singlet state.
However, this fidelity degrades as the magnetic field increases, due to leakage to the T\textsuperscript{-} state.
We then pulse various detunings in (1,1), finally coming back to the readout position, labelled R in Fig. \ref{fig:Readout}(a,c). 
When the pulse amplitude matches the detuning position of the S-T\textsuperscript{-} anticrossing, the singlet rapidly mixes with the triplet, and we observe a return singlet probability of around 50\%.  
By varying the magnetic field, and therefore the position of the S-T- anticrossing, we can trace the singlet energy branch.
Figure \ref{fig:Readout}(d) presents the results of one such experiment, where the singlet return probability is plotted as a function of both magnetic field and detuning pulse amplitude for a square pulse duration of $\SI{400}{\nano\second}$. 
This measurement allows us to extract the residual exchange interaction in (1,1) $J\tss{0} = \SI{60}{\mega\hertz}$ for a confinement potential C = 0.7 V. However, this value can be tuned by adjusting the confinement potential despite the absence of a coupling gate.

\section{Spin visibility of coherent exchange oscillations} \label{section:spin visibility}

Next, we implement a qubit on the natural basis formed by the singlet (S) and unpolarized triplet (T\tss{0}) states of our DQD, see Fig. \ref{fig:Readout}(c), \cite{Petta2005}. 
We then operate this qubit in the two-spin subspace using only gate voltages, and use gate-reflectometry to read it. 
\begin{figure*}
\includegraphics[width=\textwidth]{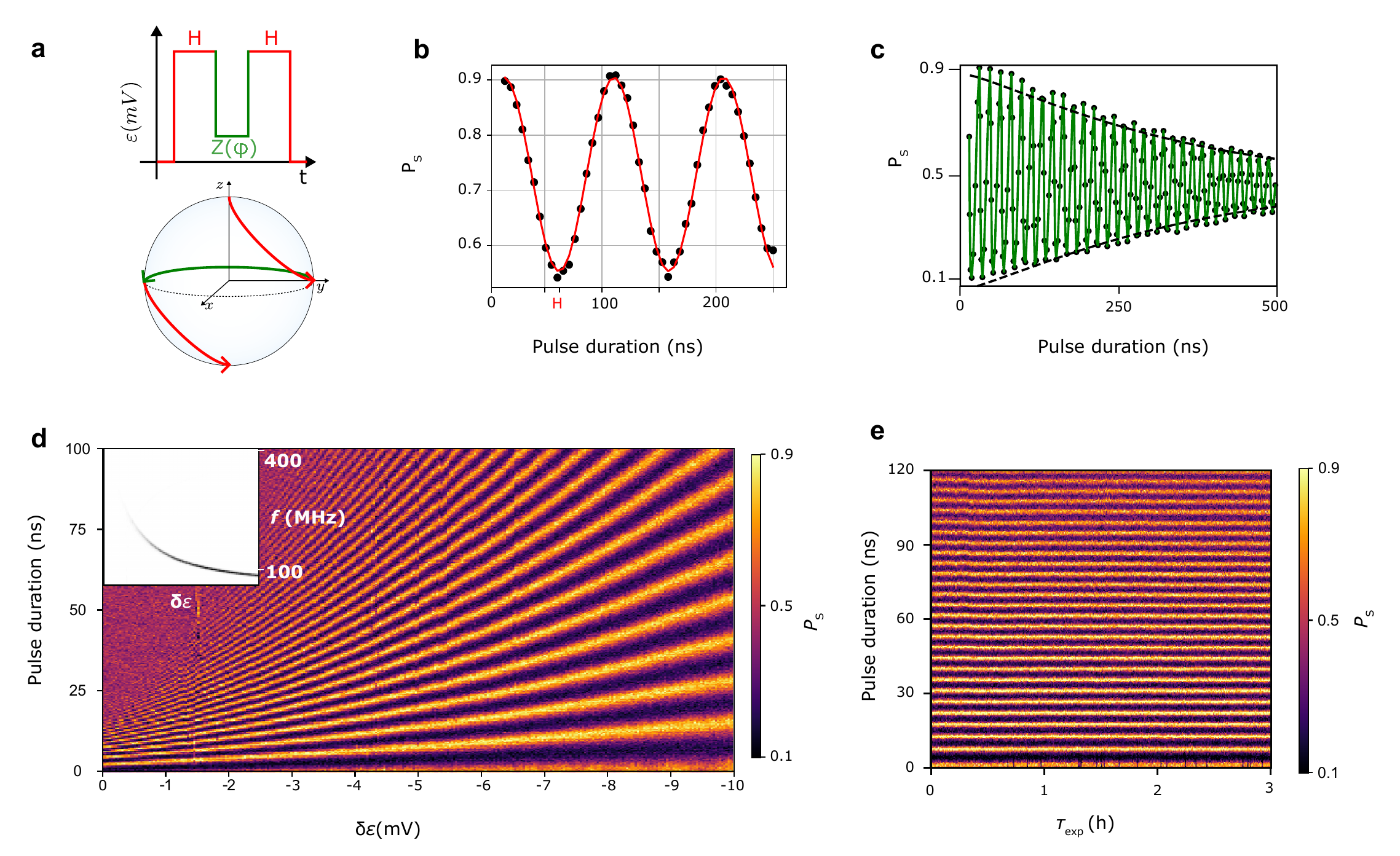}
\caption{
\textbf{Spin visibility of exchange-driven coherent oscillations at low temperature.}
\textbf{a} Pulse sequence used to measure exchange oscillation. A Hadamard gate is first applied to bring the state to the equatorial plane. The detuning is then modified to apply a phase gate. The qubit is subsequently allowed to evolve in the equatorial plane before it is projected back to the z axis by applying a second H gate. A specific case of a X = HZH gate is presented on the Bloch sphere.
\textbf{b} Oscillations performed for $J = \Delta B_z$. This curve was used to calibrate the H gate. 
\textbf{c} Exchange oscillations for $J$ = 68 MHz reveal a spin visibility of 80\%. The dashed envelop corresponds to a fit with $exp(-t/T_2^*)^\alpha$, with $\alpha=1.3$ and $T_2^*$ = 420 ns. The extracted quality factor is Q = 28. 
\textbf{d} Exchange oscillations obtained using the pulse sequence described in (a). Inset: FFT of the exchange oscillations versus detuning.
\textbf{e} Exchange oscillations recorded at $J$ = 220 MHz over 3 hours.
}
\label{fig:Exchange}%
\end{figure*}

Despite the absence of a coupling gate, we can tune the exchange interaction over one order of magnitude by playing with the confinement potential. We tune the residual exchange interaction in (1,1) around $J\tss{0} = \SI{6}{\mega\hertz}$ at C = 0.8 V, we can produce exchange oscillations in a Ramsey-like experiment with the pulse sequence described in the inset of Fig. \ref{fig:Exchange}(a). 
As before, we start by initializing a singlet state in (2,0) using the method described for the funnel experiment (point I in Fig \ref{fig:Readout} (c)). 
We then create a superposition of states by executing a Hadamard gate (H), see Fig. \ref{fig:Exchange}(b). 
The H gate is implemented by pulsing to a position where the exchange energy J is equal to $\Delta B_{z}$, the difference in Zeeman energy between the two dots $J = \Delta B\tss{z} = \SI{6}{\mega\hertz}$ at 1 T, and staying there for a time $\tau\tss{H} = 1/2\sqrt{J^2+\Delta B_{z}^2}$, see Fig. \ref{fig:Exchange}(b).
During this time, the initial singlet rotates by $\pi$ around an axis $\vec{y}+\vec{z}$ to end up in the \updown state \cite{PhysRevLett.110.146804, Connors2022}.
Applying a Z phase gate changes the phase of the qubit by pulsing to a detuning were $J >> \Delta B_z$.
We then let the system freely evolve for a time $\tau_{p}$, during which it accumulates a phase at a rate proportional to $\sqrt{J^{2} + \Delta B_{z}^{2}} \approx J$.
Applying another H gate at the end of this free evolution returns the system to the S-T\tss{0} basis, allowing its measurement during the readout step (R in  Fig \ref{fig:Readout} (d)).
The singlet probability determined with respect to the duration of the free evolution step is presented in Fig. \ref{fig:Exchange}(c). By changing the detuning at which the Z gate is performed, we can change the frequency of the exchange oscillations, see Fig. \ref{fig:Exchange}(d), producing oscillations at rates of up to 300 MHz.

On this coherent oscillation, spin visibility is around 80\% for an integration time of \SI{100}{\micro\second}. 
This corresponds to a spin readout fidelity of around 90\%, which is less than the 95\% charge readout fidelity previously obtained (Fig. \ref{fig:Readout}(b)).
This degradation may be linked to the reduction in signal due to the relatively high magnetic field used in the experiment (\SI{1}{\tesla}). In effect, the magnetic field causes a shift and a reduction in the quality factor of the resonator,  halving the signal amplitude.

We next characterize the qubit's quality in terms of decoherence induced by charge noise at low temperature.
By fitting the coherent oscillations in Fig. \ref{fig:Exchange} \cite{Keith2022}, we obtain a coherence time of 420 ns for an exchange value of 68 MHz. 
These values give a quality factor of around Q = 28. 
From this Q factor, we can roughly estimate the fidelity of a $\sqrt{\text{SWAP}}$ gate, or maximally entangled gate comprising 2 Loss-Divicenzo qubits. 
Using (F=1-t$_{g}/T^*_2$), where t$_{g}$ is the gate time \cite{Stano2022}, fidelity on the order of 99\% can therefore be expected.
Full characterization of the  fidelity of the two-qubit gate would require randomized benchmarking combined with arbitrary initialization of both spins. 
Nevertheless, we take this first estimation of fidelity as a demonstration of the high quality of the device in terms of charge fluctuations.

To extract the charge noise for the qubit unit cell more quantitatively, we probe electrical fluctuations using exchange oscillations \cite{Connors2022, PhysRevLett.110.146804, Kranz2020}.
We start by characterizing the qubit's sensitivity to noise by exploiting the dependence of the exchange energy on detuning.
By performing Fourier transformation of the oscillations shown in Fig. \ref{fig:Exchange}(c), we extract the value of the detuning-dependent exchange coupling, which follows the expected trend $J=\epsilon+\sqrt{\epsilon^2+4t_c^2}$ (see supplementary materials).
Using a sensitivity calibration curve built from the dependence of J with detuning, sensitivity values of up to d$J$/d$\epsilon$=$\SI{1}{\mega\hertz}.\mu$eV$^{-1}$ are reached.
Figure \ref{fig:Exchange}(e) presents exchange oscillations recorded over 3 hours at points of high sensitivity (1 MHz/$\mu$eV).
In the supplementary materials, we present the power spectral density (PSD) of these measurements, revealing that this parameter reaches a plateau above $10^{-3}$ Hz. 
This result highlights the limit of this method when seeking to extract very low charge noise powers. In particular, we are limited by the resolution of the FFT, which sets the limits of our noise floor to 10 $\mu$eV$^{2}$/Hz.
Nevertheless, our qualitative results reveal extremely stable oscillations, confirming the good performance of the device at 80 mK.

\section{Hot qubit operation}

\begin{figure*}%
\includegraphics[width=\textwidth]{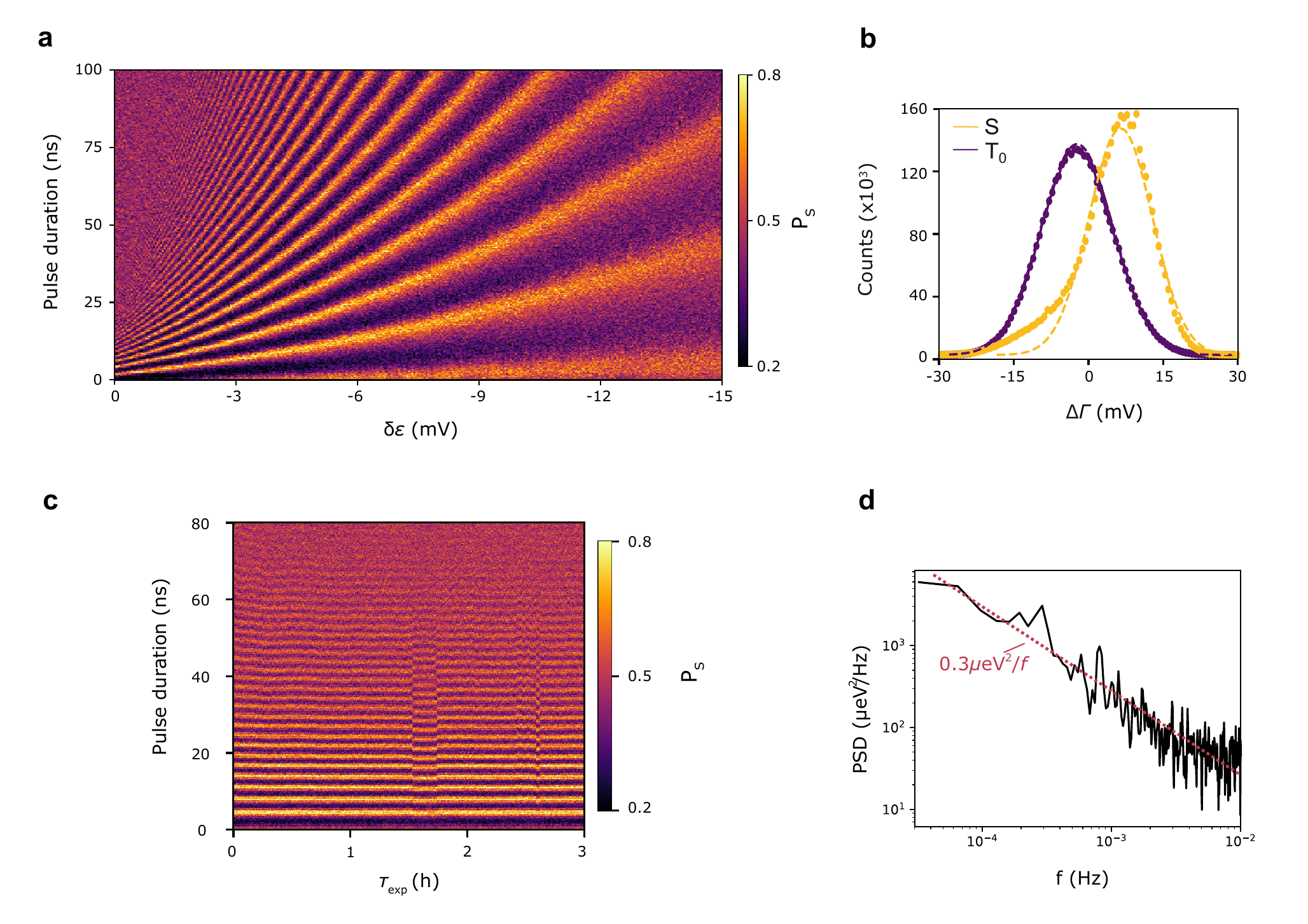}
\caption{
\textbf{Hot operation of the qubit} 
\textbf{a} Exchange oscillations at 1 K.
\textbf{b}  Histograms showing the singlet state initialized (yellow) and the triplet state (purple) produced after application of a HZH to the singlet state. Both distibutions are fitted by Gaussian curves. Relaxation during measurement of singlets is revealed by the extra population at negative signal value relative to the fitted curve. The extracted fidelity from these distributions is 75\%, in agreement with the 50\% visibility extracted in (a).
\textbf{c} Exchange oscillations recorded at $J$ = 310 MHz over 3 hours.
\textbf{d} Power spectral density of the charge noise at 1 K extracted from (c). The trend follows the red dashed line, corresponding to a fit to A/f, with A=0.3 $\mu$eV$^{2}/Hz$.
}
\label{fig:Hot}%
\end{figure*}
1-K operation of spin qubits drastically relaxes experimental constraints \cite{Yang2020, Huang2024}. In particular, it makes it possible to work in cryogenic systems with high available cooling power, a crucial milestone on the pathway toward the co-integration of classical dissipative electronic chips and quantum chips \cite{Xue2021, Bartee2024}.

We characterize our qubit unit cell at 1 K both in terms of readout and of coherent control.
Fig. \ref{fig:Hot}(a) shows the coherent exchange oscillations as a function of detuning at 1 K. Due to a reduction in readout signal amplitude relative to the measurements performed at lower temperature, spin visibility is 50\%. Figure \ref{fig:Hot}(b)  shows the distribution, revealing that only the signal amplitude has changed, not the noise contribution to SNR (width of the distributions) and is only limited by the noise temperature of the amplification chain.

To extract the charge noise for the qubit unit cell more quantitatively at 1 K, as for low temperature, we exploit exchange oscillations to probe electrical fluctuations.
Figure \ref{fig:Hot}(c) shows the exchange oscillations measured at points of high sensitivity at 1 K, recorded over 3 hours.
From these measurements, we can extract the power spectral density (PSD) (Fig. \ref{fig:Hot}(d)).

To do so, we first apply a Fourier-transform to  the time-dependent variation of the exchange energy, and normalize the result using a sensitivity to detuning noise of $\SI{1}{\mega\hertz}.\mu$eV$^{-1}$ (extracted from the derivative of $J(\epsilon)$ in supplementary materials).
We obtain a 1/f trend which confirms the electrical origin of the noise for large exchange interactions.
Moreover, the amplitude of charge noise detuning extrapolated at 1 Hz is $0.3 \mu $eV$^2$/Hz at 1 K. 
This value is similar to that obtained with other semiconductor spin qubit platforms at lower temperatures \cite{Connors2022, Jock2022}, and is two orders of magnitude smaller compared to a different qubit studied on the same platform \cite{Klemt2023}. We believe that this improvement is mainly influenced by the isolation from reservoirs; charge noise is reduced by decoupling from two-level systems located at the edges of the source and drain junctions.

\section{Conclusion}
In conclusion, here we have demonstrated the first single-shot measurement of a spin qubit using in-situ gate-reflectometry.
We show a visibility above 80\% in the coherent control of a qubit, despite its relatively small lever arm to the resonator gate (<0.15 eV.V$^{-1}$).
Exploiting this readout, we show how an isolated double quantum dot in a foundry-fabricated device can form a qubit unit cell that operates effectively at temperatures up to 1 K. This cell is compatible with both 2-electron spin coherent control and in-situ readout. The coherent control shows very good metrics: quality factor coherence time, influence of charge noise at very low temperature and, significantly, at 1 K. 

In future iterations, the qubit visibility could be increased by speeding up the measurement integration time to compensate for the short relaxation time. Signal-to-noise ratio could be further improved by increasing both the quality factor and the characteristic impedance of the resonator, as well as its stability with respect to the applied magnetic field. Finally, the ability to rapidly change the tunnel coupling (not possible in the present experiment) could help to optimize the readout configuration during the measurement sequence. 
For large-scale spin-based quantum processors, our unit cell which does not possess single-qubit functionnality could be combined with global magnetic microwave control \cite{Vahapoglu2022} to achieve universal operations. This creates a clear path for compact scale-up, with our unit cell as the basic building block.

\section{METHODS}

\subsection{Experimental setup}
Experiments were performed on the CMOS device shown in Fig. \ref{fig:Device}(a) in a dilution refrigerator with a base temperature of 80 mK. Gate-based reflectometry was achieved using analog modulation/demodulation followed by digitization using an NI ADC board. Gate voltages were applied using digital-to-analog converters controlled by an sbRIO-9208 FPGA board.

\section{ACKNOWLEDGEMENTS}
We acknowledge technical support from L. Hutin, D. Lepoittevin, I. Pheng, T. Crozes, L. Del Rey, D. Dufeu, M. Guillot, J. Jarreau, C. Hoarau and C. Guttin. 
We thank E. Chanrion, P.-A. Mortemousque, V. Champain and B. Brun-Barriere for fruitful discussions.
This work was supported by the Agence Nationale de la Recherche through the PEPR PRESQUILE project ANR-22-PETQ-0002. This project was also partly funded through the QuCube project (Grant agreement No.810504) and the QLSI2 project (Grant agreement No.101174557).

\section{AUTHORS CONTRIBUTIONS}
P.H. carried out the experiment with the help of M.T., G.H. and J. N. M.C.D. wrote the instrumental interface environment to control the setup. B.B, H. N. and M.V. designed and fabricated the device. B.M. simulated the electrostatic and exchange interaction in the device.  M.U. supervised the project with the help of T.M. 
 P.H. and M.U. wrote the manuscript with inputs from all the authors.

\section{COMPETING INTERESTS}
The authors declare no competing financial or non-financial interests

\section{DATA AVAILABILITY}
All data supporting the findings of this study are available from the corresponding author upon request.

\clearpage

\section{Appendix}
\appendix

\section{Appendix A : Results on the (0,2) to (1,1) transition}

\begin{figure*}
\includegraphics[width=12cm, height=10cm]{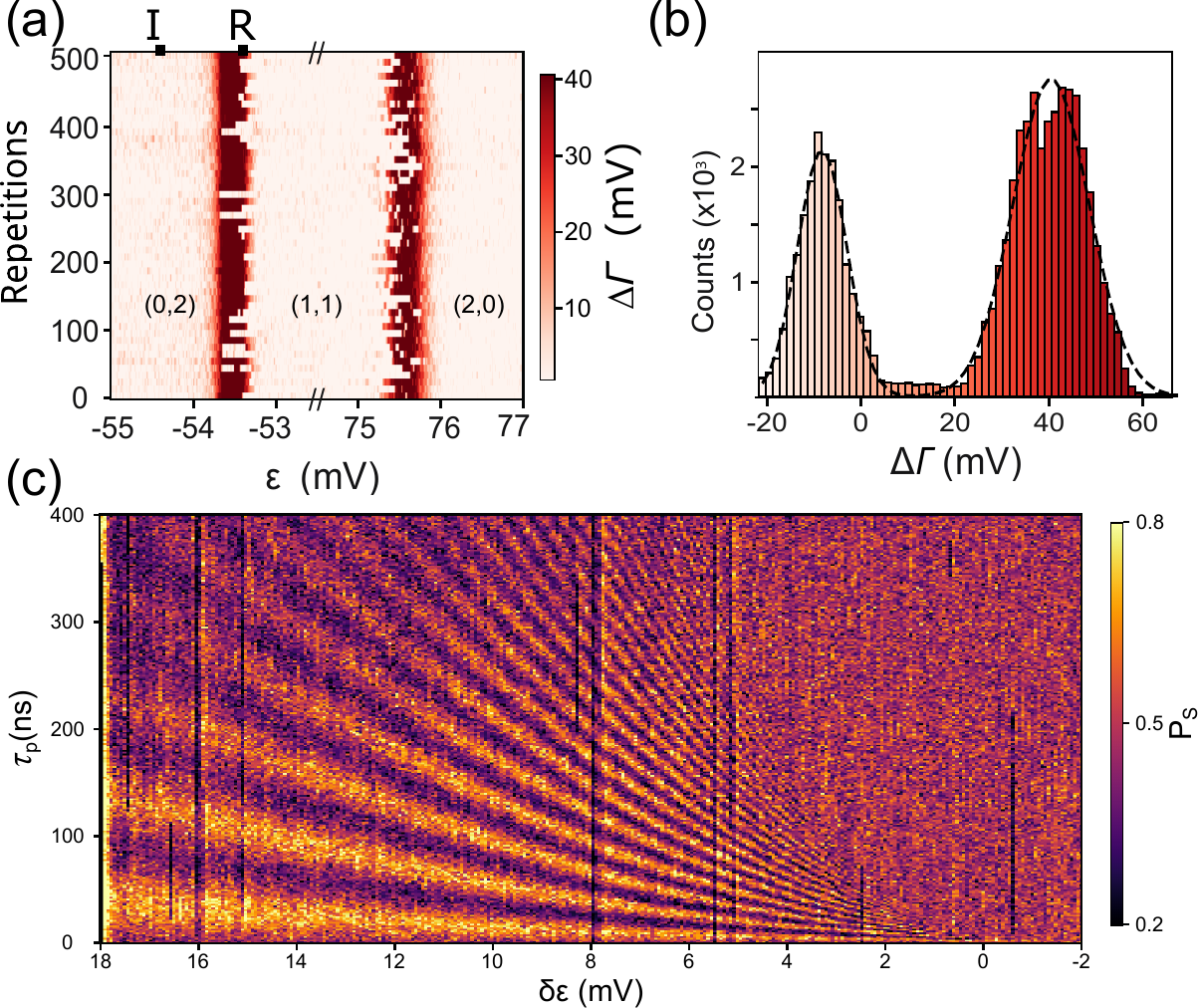}
\caption{
\textbf{Extended data for the (0,2)$\leftrightarrow$(1,1) transition} 
\textbf{(a)} Charge stability diagram showing the new readout (R) and initialization (I) positions.
\textbf{(b)} Zero-field Histogram obtained by binning the repeated signal at measurement position R on the (0,2)$\leftrightarrow$(1,1) transition to probe the singlet and triplet  populations.
\textbf{(c)} Coherent exchange oscillations at 1 T as a function of the detuning from the readout position for the (0,2)$\leftrightarrow$(1,1) transition.
}
\label{fig:upper_transi}%
\end{figure*}

In the main text, we have described results obtained at the optimal working point on the transition with the highest quality signal. 
In this section, we present the data on the second transition which is taken with a readout position at $\varepsilon$ = \SI{-53.5}{\milli\volt} more than 120 mv away from the working point of the main text at $\varepsilon$ = \SI{75.5}{\milli\volt} (Fig. \ref{fig:upper_transi} (a)).
By placing our measurement position at the point of maximum contrast, we record 10 000 samples. 
The data are illustrated in the histogram in Fig. \ref{fig:upper_transi} (b). 
We obtain at zero field a charge fidelity of 97\% for an integration time of \SI{100}{\micro\second}, similar to what we have for the first transition.
Following the protocol described in section 3, we can make coherent exchange oscillations at 1 T by applying an Hadamard gate, letting the system freely evolve under an exchange interaction $J$ large with respect to the difference in Zeeman energy between the two dots $\Delta B_z$ and projecting back to the S-T\tss{0} basis.
We present the data obtained in Fig. \ref{fig:upper_transi}, showing a visibility around 50\%.
We can conclude that transitions between (1,1) and (0,2) or (2,0) are similar in term of tunnel coupling and exchange interaction, while these quantity are exponentially sensitive with electric field and therefore Coulomb disorder.
This highlight that this foundry-fabricated devices have reach a high level of maturity in term of charge noise and disorder.

\section{Appendix B : Power spectral density and charge noise for cold measurements}
\begin{figure}
\begin{center}
\includegraphics[width=0.5\textwidth]{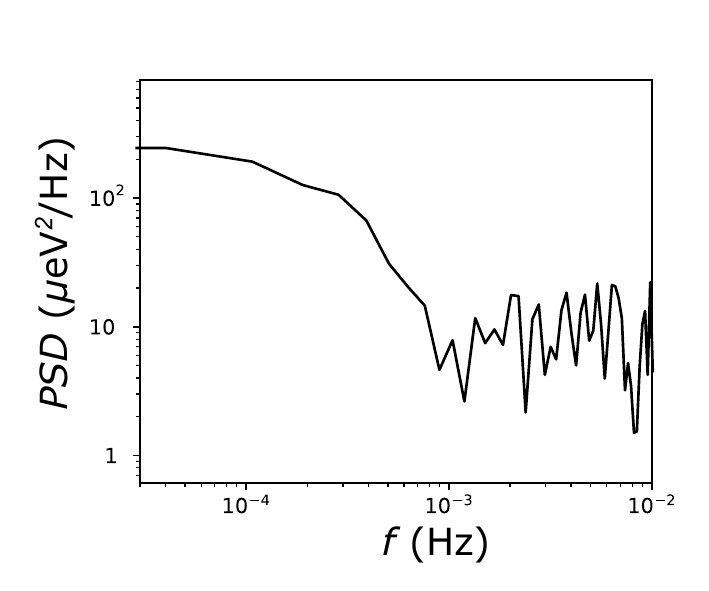}
\caption{
\textbf{Power spectral density at low temperature} 
Free induction decay measurements allows to extract the fluctuations of J over time. The error on the extraction of the FID frequency is 3MHz which leads to the plateau observed above 10$^{-3}$ Hz.
}
\label{fig:PSD_LT}%
\end{center}
\end{figure}

Similar to the 1K analysis presented in the main text, we quantitatively determined the charge noise power spectral density within the qubit unit cell at 80mK by analyzing exchange oscillations, which are sensitive indicators of electrical fluctuations. Figure \ref{fig:Exchange}(e) displays these oscillations, measured over three hours at points of peak sensitivity. From these measurements, we derived the noise's power spectral density (PSD), illustrated in Figure \ref{fig:PSD_LT}.
The sensitivity of this measurement is directly related to the limitation of the method: the Fourier transform of the time dependent oscillations gives a frequency resolution of 3 MHz. Using the sensitivity at the point of measurement we end up with a noise floor of 10 $\mu$eV$^{2}$/Hz.

\section{Appendix C : Numerical simulation of exchange interaction in the isolated double quantum dot}

In this section we provide the simulated results on the dependence of exchange on detuning shown in Fig. \ref{fig:Exchange} (d) of the main text. We performed numerical simulations on the realistic device geometry that we sample on a three-dimensional grid. 
We solved Poisson’s equation within this grid to account for the realistic device electrostatics, and fed the resulting potential to an effective mass solver to obtain the single particle energies and wave-functions. 
Since the experiment operates in the (1,1)-(2,0) charge configuration, correlation effects need to be accounted for. 
We computed the two-particle energies and wavefunctions using a full configuration-interaction (FCI) method in a constricted basis of single-particle states.
We first estimated the two-particle tunnel coupling ($\tau_{2p}$) in the experiment by fitting the dependence of $J$ on detuning to 
\begin{equation}
J =   \frac{\delta \varepsilon}{2}+ \frac{1}{2} \sqrt{ \delta \varepsilon^2 + 4 \tau_{2p} ^2}		    
\end{equation}
which arises from the diagonalization of the toy-model Hamiltonian for the S(0,2), S(1,1) and T(1,1) states \cite{PhysRevB.59.2070}. 
From the fit, we obtain $\tau_{2p} \approx 10 \mu$eV. 
In the simulations, we biased the two face-to-face gates V$_G$, V$_T$ to create a double QD, and we grounded all other gates in the device. 
In order to define a bias point in the simulations that is meaningful for comparison with the experiment, we chose the G and T gate bias to match the experimental t$_{2p} \approx 10 \mu$eV. 
This is reached for C=(V$_G$ + V$_T$)/2 = 0.674 V, which is remarkably close to the experimental values shown in the main text. 
We then scanned the exchange at the vicinities of the (1,1)-(2,0) anticrossing by performing individual FCI simulations. We plot the results, together with the experimental data, in Fig.7. There is a remarkable agreement between simulation and experiment, both in the magnitude of exchange and on the dependence on detuning. We attribute the slight mismatch in the values at small detuning to small differences in the experimental and simulated lever-arms. 
Note that the simulations are made in a pristine device, in absence of electrical disorder. Several works have reported the strong impact of disorder on the QD properties \cite{Cifuentes2024,PhysRevApplied.17.024022}, as well as on the exchange interaction \cite{PhysRevApplied.22.024030}. It is also important to point out that there are no fitting parameters in the simulation-experimental comparison. These results highlight once again the quality of the experimental device, whose electrostatics behave remarkably similar to the simulated one for a pristine device. 
\begin{figure}
\begin{center} 
\includegraphics[width=0.7\textwidth]{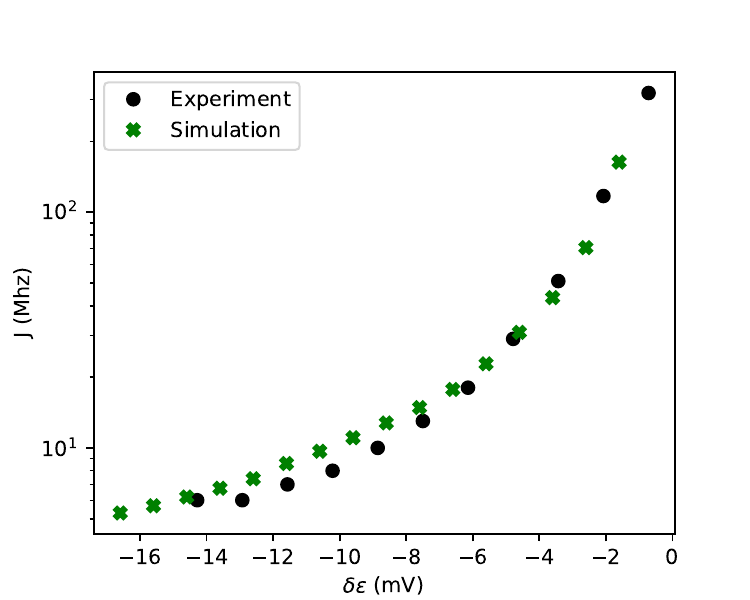}
\caption{
\textbf{Numerical simulation of exchange interaction} 
Comparison between the simulated and experimental dependence of exchange on detuning. In the simulations, C=(V$_G$ + V$_T$)/2 = 0.674 V. 
}
\end{center}
\label{fig:simu}%
\end{figure}
\clearpage

\bibliographystyle{ieeetr} 
\bibliography{biblio}

\begin{thebibliography}{10}

\bibitem{Watson2018}
T.~F. Watson, S.~G.~J. Philips, E.~Kawakami, D.~R. Ward, P.~Scarlino, M.~Veldhorst, D.~E. Savage, M.~G. Lagally, M.~Friesen, S.~N. Coppersmith, M.~A. Erikson, and L.~M.~K. Vandersypen, ``A programmable two-qubit quantum processor in silicon,'' {\em Nature}, vol.~555, p.~633, 2018.

\bibitem{Philips_2022}
S.~G.~J. Philips, M.~T. Mądzik, S.~V. Amitonov, S.~L. de~Snoo, M.~Russ, N.~Kalhor, C.~Volk, W.~I.~L. Lawrie, D.~Brousse, L.~Tryputen, B.~P. Wuetz, A.~Sammak, M.~Veldhorst, G.~Scappucci, and L.~M.~K. Vandersypen, ``Universal control of a six-qubit quantum processor in silicon,'' 2022.

\bibitem{Huang2024}
J.~Y. Huang, R.~Y. Su, W.~H. Lim, M.~Feng, B.~van Straaten, B.~Severin, W.~Gilbert, N.~Dumoulin~Stuyck, T.~Tanttu, S.~Serrano, J.~D. Cifuentes, I.~Hansen, A.~E. Seedhouse, E.~Vahapoglu, R.~C.~C. Leon, N.~V. Abrosimov, H.-J. Pohl, M.~L.~W. Thewalt, F.~E. Hudson, C.~C. Escott, N.~Ares, S.~D. Bartlett, A.~Morello, A.~Saraiva, A.~Laucht, A.~S. Dzurak, and C.~H. Yang, ``High-fidelity spin qubit operation and algorithmic initialization above 1 k,'' {\em Nature}, vol.~627, pp.~772--777, Mar 2024.

\bibitem{Takeda2022}
K.~Takeda, A.~Noiri, T.~Nakajima, T.~Kobayashi, and S.~Tarucha, ``Quantum error correction with silicon spin qubits,'' {\em Nature}, vol.~608, pp.~682--686, Aug 2022.

\bibitem{Weinstein2023}
A.~Weinstein, M.~Reed, A.~Jones, {\em et~al.}, ``Universal logic with encoded spin qubits in silicon,'' {\em Nature}, vol.~615, pp.~817--822, 2023.

\bibitem{Thorvaldson2024}
I.~Thorvaldson, D.~Poulos, C.~Moehle, {\em et~al.}, ``Grover’s algorithm in a four-qubit silicon processor above the fault-tolerant threshold,'' {\em ArXiv}, vol.~2404.08741, 2024.

\bibitem{Chatterjee2021}
A.~Chatterjee, P.~Stevenson, S.~De~Franceschi, A.~Morello, N.~P. de~Leon, and F.~Kuemmeth, ``Semiconductor qubits in practice,'' {\em Nature Reviews Physics}, vol.~3, pp.~157--177, Mar 2021.

\bibitem{Zwerver2022}
A.~Zwerver, T.~Krähenmann, T.~Watson, {\em et~al.}, ``Qubits made by advanced semiconductor manufacturing,'' {\em Nature Electronics}, vol.~5, pp.~184--190, 2022.

\bibitem{Crawford2023}
O.~Crawford, J.~Cruise, N.~Mertig, and M.~Gonzalez-Zalba, ``Compilation and scaling strategies for a silicon quantum processor with sparse two-dimensional connectivity,'' {\em npj Quantum Information}, vol.~9, p.~13, 2023.

\bibitem{Neyens2024}
S.~Neyens, O.~Zietz, T.~Watson, {\em et~al.}, ``Probing single electrons across 300-mm spin qubit wafers,'' {\em Nature}, vol.~629, p.~80, 2024.

\bibitem{Thomas2025}
E.~J. Thomas, V.~N. Ciriano-Tejel, D.~F. Wise, D.~Prete, M.~d. Kruijf, D.~J. Ibberson, G.~M. Noah, A.~Gomez-Saiz, M.~F. Gonzalez-Zalba, M.~A.~I. Johnson, and J.~J.~L. Morton, ``Rapid cryogenic characterization of 1,024 integrated silicon quantum dot devices,'' {\em Nature Electronics}, vol.~8, pp.~75--83, Jan 2025.

\bibitem{10639218}
N.~D. Stuyck, M.~Feng, W.~H. Lim, S.~S. Ramirez, C.~C. Escott, T.~Botzem, T.~Tanttu, C.~H. Yang, A.~Saraiva, A.~Laucht, S.~Kubicek, J.~Jussot, S.~Beyne, B.~Raes, R.~Li, C.~Godfrin, D.~Wan, K.~De~Greve, and A.~S. Dzurak, ``Demonstration of 99.9

\bibitem{Steinacker2024}
P.~Steinacker, N.~Dumoulin~Stuyck, W.~H. Lim, and et~al., ``A 300 mm foundry silicon spin qubit unit cell exceeding 99\% fidelity in all operations,'' {\em arXiv}, p.~2410.15590, 2024.
\newblock https://arxiv.org/abs/2410.15590.

\bibitem{Veldhorst2017}
M.~Veldhorst, H.~G.~J. Eenink, C.~H. Yang, and A.~S. Dzurak, ``Silicon cmos architecture for a spin-based quantum computer,'' {\em Nat. Commun.}, vol.~8, p.~1766, 2017.

\bibitem{jung2012radio}
M.~Jung, M.~D. Schroer, K.~D. Petersson, and J.~R. Petta, ``Radio frequency charge sensing in inas nanowire double quantum dots,'' {\em Applied Physics Letters}, vol.~100, no.~25, p.~253508, 2012.

\bibitem{house2016high}
M.~G. House, T.~Kobayashi, B.~Weber, S.~J. Hile, T.~F. Watson, J.~van~der Heijden, S.~Rogge, and M.~Y. Simmons, ``High-sensitivity charge detection with a radio-frequency quantum point contact,'' {\em Physical Review Applied}, vol.~6, no.~4, p.~044016, 2016.

\bibitem{hornibrook2014frequency}
J.~M. Hornibrook, J.~I. Colless, A.~C. Mahoney, X.~G. Croot, S.~Blanvillain, H.~Lu, A.~C. Gossard, and D.~J. Reilly, ``Frequency multiplexing for readout of spin qubits,'' {\em Applied Physics Letters}, vol.~104, no.~10, p.~103108, 2014.

\bibitem{Oakes_2022}
G.~A. Oakes, V.~N. Ciriano-Tejel, D.~F. Wise, M.~A. Fogarty, T.~Lundberg, C.~Lain\'e, S.~Schaal, F.~Martins, D.~J. Ibberson, L.~Hutin, B.~Bertrand, N.~Stelmashenko, J.~W.~A. Robinson, L.~Ibberson, A.~Hashim, I.~Siddiqi, A.~Lee, M.~Vinet, C.~G. Smith, J.~J.~L. Morton, and M.~F. Gonzalez-Zalba, ``Fast high-fidelity single-shot readout of spins in silicon using a single-electron box,'' {\em Phys. Rev. X}, vol.~13, p.~011023, Feb 2023.

\bibitem{PhysRevX.5.031024}
M.~Urdampilleta, A.~Chatterjee, C.~C. Lo, T.~Kobayashi, J.~Mansir, S.~Barraud, A.~C. Betz, S.~Rogge, M.~F. Gonzalez-Zalba, and J.~J.~L. Morton, ``Charge dynamics and spin blockade in a hybrid double quantum dot in silicon,'' {\em Phys. Rev. X}, vol.~5, p.~031024, Aug 2015.

\bibitem{gonzalez2015gate}
M.~F. Gonzalez-Zalba, S.~Barraud, A.~J. Ferguson, and A.~C. Betz, ``Gate-based high fidelity spin readout in a cmos device,'' {\em Nature Communications}, vol.~6, p.~6084, 2015.

\bibitem{crippa2019}
A.~Crippa, R.~Ezzouch, A.~Aprá, A.~Amisse, R.~Laviéville, L.~Hutin, B.~Bertrand, M.~Vinet, M.~Urdampilleta, T.~Meunier, M.~Sanquer, X.~Jehl, R.~Maurand, and S.~D. Franceschi, ``Gate-reflectometry dispersive readout and coherent control of a spin qubit in silicon,'' {\em Nature Communications}, vol.~10, p.~2776, 2019.

\bibitem{Zheng2019}
G.~Zheng, N.~Samkharadze, M.~L. Noordam, {\em et~al.}, ``Rapid gate-based spin read-out in silicon using an on-chip resonator,'' {\em Nature Nanotechnology}, vol.~14, p.~742, 2019.

\bibitem{Pakkiam2018}
P.~Pakkiam, A.~Timofeev, M.~House, M.~Hogg, T.~Kobayashi, M.~Koch, S.~Rogge, and M.~Simmons, ``Single-shot single-gate rf spin readout in silicon,'' {\em Physical Review X}, vol.~8, p.~041032, 2018.

\bibitem{DzurakReflecto}
A.~West, B.~Hensen, A.~Jouan, T.~Tanttu, C.~Yang, A.~Rossi, M.~Gonzalez-Zalba, F.~Hudson, A.~Morello, D.~Reilly, and A.~Dzurak, ``Gate-based single-shot readout of spins in silicon,'' {\em Nature Nanotechnology}, vol.~14, p.~437, 2019.

\bibitem{Ruffino_2021}
A.~Ruffino, T.-Y. Yang, J.~Michniewicz, Y.~Peng, E.~Charbon, and M.~F. Gonzalez-Zalba, ``A cryo-{CMOS} chip that integrates silicon quantum dots and multiplexed dispersive readout electronics,'' {\em Nature Electronics}, vol.~5, pp.~53--59, dec 2021.

\bibitem{Xue_2021}
X.~Xue, B.~Patra, J.~P.~G. van Dijk, N.~Samkharadze, S.~Subramanian, A.~Corna, B.~Paquelet~Wuetz, C.~Jeon, F.~Sheikh, E.~Juarez-Hernandez, B.~P. Esparza, H.~Rampurawala, B.~Carlton, S.~Ravikumar, C.~Nieva, S.~Kim, H.-J. Lee, A.~Sammak, G.~Scappucci, M.~Veldhorst, F.~Sebastiano, M.~Babaie, S.~Pellerano, E.~Charbon, and L.~M.~K. Vandersypen, ``Cmos-based cryogenic control of silicon quantum circuits,'' {\em Nature}, vol.~593, no.~7858, pp.~205--210, 2021.

\bibitem{Bartee2024}
S.~Bartee, W.~Gilbert, K.~Zuo, {\em et~al.}, ``Spin qubits with scalable milli-kelvin cmos control,'' {\em ArXiv}, vol.~2407.15151, 2024.

\bibitem{Hamonic2024}
P.~Hamonic, M.~Nurizzo, J.~Nath, M.~C. Dartiailh, V.~El-Homsy, M.~Fragnol, B.~Martinez, P.-L. Julliard, B.~C. Paz, M.~Ouvrier-Buffet, J.-B. Filippini, B.~Bertrand, H.~Niebojewski, C.~Bäuerle, M.~Vinet, F.~Balestro, T.~Meunier, and M.~Urdampilleta, ``Combining multiplexed gate-based readout and isolated {CMOS} quantum dot arrays,'' Oct. 2024.
\newblock arXiv:2410.02325 [cond-mat].

\bibitem{PRXQuantum.3.040335}
D.~Niegemann {\em et~al.}, ``Parity and singlet-triplet high-fidelity readout in a silicon double quantum dot at 0.5 k,'' {\em PRX Quantum}, vol.~3, p.~040335, 2022.

\bibitem{Klemt2023}
B.~Klemt, V.~El-Homsy, M.~Nurizzo, P.~Hamonic, {\em et~al.}, ``Electrical manipulation of a single electron spin in cmos with micromagnet and spin-valley coupling,'' {\em npj Quantum Information}, vol.~9, p.~107, 2023.

\bibitem{PhysRevApplied.14.024066}
E.~Chanrion {\em et~al.}, ``Charge detection in an array of cmos quantum dots,'' {\em Physical Review Applied}, vol.~14, p.~024066, 2020.

\bibitem{Bertrand2015}
B.~Bertrand, H.~Flentje, S.~Takada, M.~Yamamoto, S.~Tarucha, A.~Ludwig, A.~D. Wieck, C.~B{\"a}uerle, and T.~Meunier, ``Quantum manipulation of two-electron spin states in isolated double quantum dots,'' {\em Physical Review Letters}, vol.~115, p.~096801, 2015.

\bibitem{ivlev2025operating}
A.~Ivlev, D.~Crielaard, M.~Meyer, W.~Lawrie, N.~Hendrickx, A.~Sammak, G.~Scappucci, C.~D{\'e}prez, and M.~Veldhorst, ``Operating semiconductor qubits without individual barrier gates,'' {\em arXiv preprint arXiv:2501.03033}, 2025.

\bibitem{Vigneau2023}
F.~Vigneau, F.~Fedele, A.~Chatterjee, {\em et~al.}, ``Probing quantum devices with radio-frequency reflectometry,'' {\em Applied Physics Reviews}, vol.~10, no.~2, 2023.

\bibitem{Fogarty_2018}
M.~A. Fogarty, K.~W. Chan, B.~Hensen, W.~Huang, T.~Tanttu, C.~H. Yang, A.~Laucht, M.~Veldhorst, F.~E. Hudson, K.~M. Itoh, D.~Culcer, T.~D. Ladd, A.~Morello, and A.~S. Dzurak, ``Integrated silicon qubit platform with single-spin addressability, exchange control and single-shot singlet-triplet readout,'' {\em Nature Communications}, vol.~9, oct 2018.

\bibitem{PRXQuantum.4.010329}
M.~Nurizzo, B.~Jadot, P.-A. Mortemousque, V.~Thiney, E.~Chanrion, D.~Niegemann, M.~Dartiailh, A.~Ludwig, A.~D. Wieck, C.~B\"auerle, M.~Urdampilleta, and T.~Meunier, ``Complete readout of two-electron spin states in a double quantum dot,'' {\em PRX Quantum}, vol.~4, p.~010329, Mar 2023.

\bibitem{PhysRevX.9.021028}
T.~Tanttu, B.~Hensen, K.~W. Chan, C.~H. Yang, W.~W. Huang, M.~Fogarty, F.~Hudson, K.~Itoh, D.~Culcer, A.~Laucht, A.~Morello, and A.~Dzurak, ``Controlling spin-orbit interactions in silicon quantum dots using magnetic field direction,'' {\em Phys. Rev. X}, vol.~9, p.~021028, May 2019.

\bibitem{Petta2005}
J.~R. Petta, A.~C. Johnson, J.~M. Taylor, E.~A. Laird, A.~Yacoby, M.~D. Lukin, C.~M. Marcus, M.~P. Hanson, and A.~C. Gossard, ``Coherent manipulation of coupled electron spins in semiconductor quantum dots,'' {\em Science}, vol.~309, p.~2180, 2005.

\bibitem{PhysRevLett.110.146804}
O.~E. Dial, M.~D. Shulman, S.~P. Harvey, H.~Bluhm, V.~Umansky, and A.~Yacoby, ``Charge noise spectroscopy using coherent exchange oscillations in a singlet-triplet qubit,'' {\em Phys. Rev. Lett.}, vol.~110, p.~146804, Apr 2013.

\bibitem{Connors2022}
E.~Connors, J.~Nelson, L.~Edge, {\em et~al.}, ``Charge-noise spectroscopy of si/sige quantum dots via dynamically-decoupled exchange oscillations,'' {\em Nature Communications}, vol.~13, p.~940, 2022.

\bibitem{Keith2022}
D.~Keith, S.~K. Gorman, Y.~He, L.~Kranz, and M.~Y. Simmons, ``Impact of charge noise on electron exchange interactions in semiconductors,'' {\em npj Quantum Information}, vol.~8, p.~17, Feb 2022.

\bibitem{Stano2022}
P.~Stano and D.~Loss, ``Review of performance metrics of spin qubits in gated semiconducting nanostructures,'' {\em Nature Reviews Physics}, vol.~4, pp.~672--688, Oct 2022.

\bibitem{Kranz2020}
L.~Kranz {\em et~al.}, ``Exploiting a single-crystal environment to minimize the charge noise on qubits in silicon,'' {\em Advanced Materials}, vol.~32, p.~2003361, 2020.

\bibitem{Yang2020}
C.~Yang, R.~Leon, J.~Hwang, {\em et~al.}, ``Operation of a silicon quantum processor unit cell above one kelvin,'' {\em Nature}, vol.~580, pp.~350--354, 2020.

\bibitem{Xue2021}
X.~Xue, B.~Patra, J.~P.~G. van Dijk, N.~Samkharadze, S.~Subramanian, A.~Corna, B.~Paquelet~Wuetz, C.~Jeon, F.~Sheikh, E.~Juarez-Hernandez, B.~P. Esparza, H.~Rampurawala, B.~Carlton, S.~Ravikumar, C.~Nieva, S.~Kim, H.-J. Lee, A.~Sammak, G.~Scappucci, M.~Veldhorst, F.~Sebastiano, M.~Babaie, S.~Pellerano, E.~Charbon, and L.~M.~K. Vandersypen, ``Cmos-based cryogenic control of silicon quantum circuits,'' {\em Nature}, vol.~593, pp.~205--210, May 2021.

\bibitem{Jock2022}
R.~M. Jock, N.~T. Jacobson, M.~Rudolph, D.~R. Ward, M.~S. Carroll, and D.~R. Luhman, ``A silicon singlet--triplet qubit driven by spin-valley coupling,'' {\em Nature Communications}, vol.~13, p.~641, Feb 2022.

\bibitem{Vahapoglu2022}
E.~Vahapoglu, J.~P. Slack-Smith, R.~C.~C. Leon, W.~H. Lim, F.~E. Hudson, T.~Day, J.~D. Cifuentes, T.~Tanttu, C.~H. Yang, A.~Saraiva, N.~V. Abrosimov, H.-J. Pohl, M.~L.~W. Thewalt, A.~Laucht, A.~S. Dzurak, and J.~J. Pla, ``Coherent control of electron spin qubits in silicon using a global field,'' {\em npj Quantum Information}, vol.~8, p.~126, Nov 2022.

\bibitem{PhysRevB.59.2070}
G.~Burkard, D.~Loss, and D.~P. DiVincenzo, ``Coupled quantum dots as quantum gates,'' {\em Phys. Rev. B}, vol.~59, pp.~2070--2078, Jan 1999.

\bibitem{Cifuentes2024}
J.~D. Cifuentes, T.~Tanttu, W.~Gilbert, J.~Y. Huang, E.~Vahapoglu, R.~C.~C. Leon, S.~Serrano, D.~Otter, D.~Dunmore, P.~Y. Mai, F.~Schlattner, M.~Feng, K.~Itoh, N.~Abrosimov, H.-J. Pohl, M.~Thewalt, A.~Laucht, C.~H. Yang, C.~C. Escott, W.~H. Lim, F.~E. Hudson, R.~Rahman, A.~S. Dzurak, and A.~Saraiva, ``Bounds to electron spin qubit variability for scalable cmos architectures,'' {\em Nature Communications}, vol.~15, p.~4299, May 2024.

\bibitem{PhysRevApplied.17.024022}
B.~Martinez and Y.-M. Niquet, ``Variability of electron and hole spin qubits due to interface roughness and charge traps,'' {\em Phys. Rev. Appl.}, vol.~17, p.~024022, Feb 2022.

\bibitem{PhysRevApplied.22.024030}
B.~Martinez, S.~de~Franceschi, and Y.-M. Niquet, ``Mitigating variability in epitaxial-heterostructure-based spin-qubit devices by optimizing gate layout,'' {\em Phys. Rev. Appl.}, vol.~22, p.~024030, Aug 2024.

\end{thebibliography}


\begin{thebibliography}{99}

\bibitem{Chaterjee2020}Chatterjee, A., Stevenson, P., De Franceschi, S. et al. Semiconductor qubits in practice.. {\em Nat. Rev. Phys.} \textbf{3}, 157–177 (2021).

\bibitem{Steinacker2024}Paul Steinacker, Nard Dumoulin Stuyck, Wee Han Lim et al. A 300 mm foundry silicon spin qubit unit cell exceeding 99\% fidelity in all operations. {\em Arxiv} \textbf{}, 2410.15590 (2024).


\bibitem{Steinacker2024}Paul Steinacker, Nard Dumoulin Stuyck, Wee Han Lim et al. A 300 mm foundry silicon spin qubit unit cell exceeding 99\% fidelity in all operations. {\em Arxiv} \textbf{}, 2410.15590 (2024).

\bibitem{Gonzalez-Zalba2021}Gonzalez-Zalba, M., Franceschi, S., Charbon, E., Meunier, T., Vinet, M. \& Dzurak, A. Scaling silicon-based quantum computing using CMOS technology. {\em Nat. Electron.} \textbf{4}, 872-884 (2021).

\bibitem{Veldhorst2017}Veldhorst, M., Eenink, H., Yang, C. \& Dzurak, A. Silicon CMOS architecture for a spin-based quantum computer. {\em Nat. Commun.} \textbf{8}, 1766 (2017).

\bibitem{PhysRevApplied.6.054013}D.~M. Zajac, T.~M. Hazard, X.~Mi, E.~Nielsen, and J.~R. Petta.\newblock Scalable gate architecture for a one-dimensional array of semiconductor spin qubits.\newblock {Phys. Rev. Applied}, \textbf{6}, 054013 (2016).

\bibitem{Patomaki2024}Patomäki, S.M., Gonzalez-Zalba, M.F., Fogarty, M.A. et al. \newblock Pipeline quantum processor architecture for silicon spin qubits..\newblock {npj Quantum Inf}, \textbf{10}, 31 (2024).

\bibitem{Crawford2023}Crawford, O., Cruise, J.R., Mertig, N. and Gonzalez-Zalba. \newblock Compilation and scaling strategies for a silicon quantum processor with sparse two-dimensional connectivity.\newblock {npj Quantum Inf}, \textbf{9}, 13 (2023).

\bibitem{Chittock-Wood2024}Chittock-Wood J., et al.\newblock Exchange control in a MOS double quantum dot made using a 300 mm wafer process, arXiv:2408.01241

\bibitem{PRXQuantum.2.010353}Ciriano-Tejel, V. et al. Spin Readout of a CMOS Quantum Dot by Gate Reflectometry and Spin-Dependent Tunneling. {\em PRX Quantum}. \textbf{2}, 010353 (2021,3).

\bibitem{PRXQuantum.3.040335}Niegemann, D. et al. Parity and Singlet-Triplet High-Fidelity Readout in a Silicon Double Quantum Dot at 0.5 K. {\em PRX Quantum}. \textbf{3}, 040335 (2022,12).

\bibitem{PhysRevApplied.14.024066}Chanrion, E. et al. Charge Detection in an Array of CMOS Quantum Dots. {\em Phys. Rev. Appl.} \textbf{14}, 024066 (2020,8).

\bibitem{Gilbert2020}Gilbert, W. et al. Single-Electron Operation of a Silicon-CMOS 2 × 2 Quantum Dot Array with Integrated Charge Sensing. {\em Nano Lett.} \textbf{20}, 7882-7888 (2020,11).

\bibitem{Ansaloni2020}Ansaloni, F., Chatterjee, A., Bohuslavskyi, H., Bertrand, B., Hutin, L., Vinet, M. \& Kuemmeth, F. Single-electron operations in a foundry-fabricated array of quantum dots. {\em Nat. Commun.} \textbf{11}, 6399 (2020).

\bibitem{Zwerver2022}Zwerver, A.M.J., Krähenmann, T., Watson, T.F. et al., M.  Qubits made by advanced semiconductor manufacturing. {\em Nat. Electron}. \textbf{5}, 184-190 (2022).



\bibitem{Yu2023} Yu, Cécile X. et al. Strong coupling between a photon and a hole spin in silicon. {\em Nat. Nanotechnol.}, 1-6 (2023).


\bibitem{Pioro-Ladriere2008}Pioro-Ladrière, M. et al. Electrically driven single-electron spin resonance in a slanting Zeeman field. {\em Nat. Phys.} \textbf{4}, 776-779 (2008).


\bibitem{Yoneda2018}Yoneda, J. et al. A quantum-dot spin qubit with coherence limited by charge noise and fidelity higher than 99.9$\%$. {\em Nat. Nanotechnol}. \textbf{13}, 102-106 (2018).

\bibitem{Kawakami2014}Kawakami, E. et al. Electrical control of a long-lived spin qubit in a Si/SiGe quantum dot. {\em Nat. Nanotechnol}. \textbf{9}, 666-670 (2014).

\bibitem{Leon2020}Leon, R. et al. Coherent spin control of s-, p-, d- and f-electrons in a silicon quantum dot. {\em Nat. Commun.} \textbf{11}, 797 (2020).

\bibitem{PhysRevApplied.11.061006}Sigillito, A., Loy, J., Zajac, D., Gullans, M., Edge, L. \& Petta, J. Site-Selective Quantum Control in an Isotopically Enriched $^{28}\text{Si}/\text{Si}_{0.7}\text{Ge}_ {0.3}$ Quadruple Quantum Dot. {\em Phys. Rev. Appl.} \textbf{11}, 061006 (2019,6).

\bibitem{PhysRevApplied.15.044042}Zhang, X. et al. Controlling Synthetic Spin-Orbit Coupling in a Silicon Quantum Dot with Magnetic Field. {\em Phys. Rev. Appl.}. \textbf{15}, 044042 (2021,4).

\bibitem{9830352}De Franceschi, S. et al. SOI technology for quantum information processing. In 2016 IEEE International Electron Devices Meeting (IEDM) (pp. 13-4). IEEE (2016).

\bibitem{Philips2022}Philips, S. et al. Universal control of a six-qubit quantum processor in silicon. {\em Nature}. \textbf{609}, 919-924 (2022).

\bibitem{Culcer2009}Culcer, D., Hu, X. \& Das Sarma, S. Dephasing of Si spin qubits due to charge noise. {\em Appl. Phys. Lett}. \textbf{95}, 73102 (2009,8).

\bibitem{Struck2020}Struck, T. et al. Low-frequency spin qubit energy splitting noise in highly purified 28Si/SiGe. {\em Npj Quantum Inf}. \textbf{6}, 40 (2020).


\bibitem{PhysRevApplied.17.034047}Spence, C. et al. Spin-Valley Coupling Anisotropy and Noise in CMOS Quantum Dots. {\em Phys. Rev. Appl.}. \textbf{17}, 034047 (2022,3).

\bibitem{Elzerman2004}Elzerman, J., Hanson, R., Willems van Beveren, L., Witkamp, B., Vandersypen, L. \& Kouwenhoven, L. Single-shot read-out of an individual electron spin in a quantum dot. {\em Nature}. \textbf{430}, 431-435 (2004).


\bibitem{Morello2010}Morello, A. et al. Single-shot readout of an electron spin in silicon. {\em Nature}. \textbf{467}, 687-691 (2010).


\bibitem{Keith2019}Keith, D. et al. Benchmarking high fidelity single-shot readout of semiconductor qubits. {\em New J. Phys}. \textbf{21}, 63011 (2019).

\bibitem{PhysRevLett.124.257701}Zhang, X. et al. Giant Anisotropy of Spin Relaxation and Spin-Valley Mixing in a Silicon Quantum Dot. {\em Phys. Rev. Lett.}. \textbf{124}, 257701 (2020,6).

\bibitem{Petit2018}Petit, L., et al. Spin lifetime and charge noise in hot silicon quantum dot qubits.{\em Phys. Rev. Lett.} 121.7: 076801 (2018).

\bibitem{Yang2013}Yang, C. et al. Spin-valley lifetimes in a silicon quantum dot with tunable valley splitting. {\em Nat. Commun.}. \textbf{4}, 2069 (2013).

\bibitem{PhysRevB.90.235315}Huang, P., \& Hu, X. Spin relaxation in a Si quantum dot due to spin-valley mixing. {\em Phys. Rev. B.} 90(23), 235315 (2014).

\bibitem{Pla2012}Pla, J. et al. A single-atom electron spin qubit in silicon. {\em Nature}, 489(7417), 541-545 (2012).


\bibitem{PhysRevLett.99.106803}Koppens, F. et al. Universal Phase Shift and Nonexponential Decay of Driven Single-Spin Oscillations. {\em Phys. Rev. Lett.}. \textbf{99}, 106803 (2007,9).


\bibitem{Takeda2023}Takeda, K. et al. A fault-tolerant addressable spin qubit in a natural silicon quantum dot. {\em Sci. Adv}. 2(8), e1600694 (2016).

\bibitem{Martinez2022}Martinez, Biel and Niquet, Yann-Michel Variability of Electron and Hole Spin Qubits Due to Interface Roughness and Charge Traps. {\em Phys. Rev. Appl.}. \textbf{24}, 024022 (2022).


\bibitem{PhysRevB.95.075403}Huang, W., Veldhorst, M., Zimmerman, N., Dzurak, A. \& Culcer, D. Electrically driven spin qubit based on valley mixing. {\em Phys. Rev. B}. \textbf{95}, 075403 (2017,2).


\bibitem{Corna2018}Corna, A. et al. Electrically driven electron spin resonance mediated by spin–valley–orbit coupling in a silicon quantum dot. {\em Npj Quantum Inf.}. \textbf{4}, 6 (2018).

\bibitem{Hao2014} Hao, X. et al. Electron spin resonance and spin–valley physics in a silicon double quantum dot. {\em Nat. Commun.} 5, 3860 (2014).

\bibitem{Huang2021}Huang, P. \& Hu, X. Fast spin-valley-based quantum gates in Si with micromagnets. {\em Npj Quantum Inf}. \textbf{7}, 162 (2021).

\bibitem{PhysRevB.97.155433}Bourdet, L. \& Niquet, Y. All-electrical manipulation of silicon spin qubits with tunable spin-valley mixing. {\em Phys. Rev. B}. \textbf{97}, 155433 (2018,4).

\bibitem{PhysRev.94.630}Carr, H. \& Purcell, E. Effects of Diffusion on Free Precession in Nuclear Magnetic Resonance Experiments. {\em Phys. Rev.}. \textbf{94}, 630-638 (1954,5).


\bibitem{Meiboom1958}Meiboom, S. \& Gill, D. Modified Spin‐Echo Method for Measuring Nuclear Relaxation Times. {\em Rev. Sci. Instrum.} \textbf{29}, 688-691 (1958,8).

\bibitem{PhysRevA.58.2733}Viola, L. \& Lloyd, S. Dynamical suppression of decoherence in two-state quantum systems. {\em Phys. Rev. A}. \textbf{58}, 2733-2744 (1998,10).

\bibitem{kawakami2016}Kawakami, E.et al. Gate fidelity and coherence of an electron spin in an Si/SiGe quantum dot with micromagnet. {\em PNAS}, 113(42), 11738-11743 (2016).

\bibitem{Connors2022}Connors, E.J., Nelson, J., Edge, L.F. et al. Charge-noise spectroscopy of Si/SiGe quantum dots via dynamically-decoupled exchange oscillations. {\em Nat. Commun.}. \textbf{13}, 940 (2022).

\bibitem{Kranz2020}Kranz, L. et al. Exploiting a Single-Crystal Environment to Minimize the Charge Noise on Qubits in Silicon. {\em Adv. Mater}. \textbf{32}, 2003361 (2020).

\end{thebibliography}

\newpage 
\clearpage

\end{document}